# Imry-Ma disordered state induced by defects of "random local anisotropy" type in the system with *O(n)* symmetry

*A.A. Berzin[1], A.I. Morosov[2]\*, and A.S. Sigov[1]*

[1] Moscow Technological University (MIREA), 78 Vernadskiy Ave., 119454 Moscow, Russian Federation
[2] Moscow Institute of Physics and Technology (State University), 9 Institutskiy per., 141700 Dolgoprudny, Moscow Region, Russian Federation

## Abstract

For the system with the *n*-component order parameter (*O(n)*-model), conditions for initiation of the Imry-Ma disordered state resulting from the influence of defects of the "random local anisotropy" type were discovered. The initiation of such a state was shown to be possible if the distribution of local anisotropy axes directions in the order parameter space is nearly isotropic, and the limiting degree of the distribution anisotropy was found. For a higher anisotropy in the distribution of local axes directions, the long-range order in the system holds true even in the presence of defects of the given type.

---

\* E-mail: mor-alexandr@yandex.ru

## 1. Introduction

In their classical paper [1] Imry and Ma came to a conclusion that in space dimensions $d < 4$ the introduction of an arbitrarily small concentration of defects of the "random local field" type into a system with continuous symmetry of the $n$-component vector order parameter ($O(n)$ model) leads to the long-range order collapse and to the occurrence of a disordered state, which in what follows will be designated as the Imry-Ma state.

Later on their arguments were extended to defects of the "random local anisotropy" type [2, 3].

On a consideration of the amorphous magnetic with similar defects near a phase transition [4] the statement was formulated that the disorder isotropy is an exacting requirement for the Imry-Ma disordered state occurrence.

It was shown as well [5] that anisotropic distribution of random local field directions in the order parameter space gives rise to the effective anisotropy quadratic in the random field value, and the corresponding anisotropy constant was evaluated. If the magnitude of this constant exceeds the critical value obtained in Ref. [6], then the Imry-Ma disordered state ceases to be energetically favorable and the long-range order is not destroyed.

In the case of two-dimensional $O(n)$-systems, it is just the effective anisotropy existence that can explain [7] the long-range order initiation (the ordering induced by random fields) [8,9].

The goal of the present paper is to conduct analogous studies into the systems with defects of the "random local anisotropy" type with the aim of formulating the conditions for the occurrence of the Imry-Ma disordered state depending on the degree of anisotropy of the distribution of local anisotropy easy axes directions in the order parameter space.

## 2. Energy of a system of classical spins

The exchange interaction energy of $n$-component localized spins $\boldsymbol{S}_i$ comprising the $d$-dimensional lattice has the form

$$W_{ex} = -\sum_{i,j>i} J_{ij} \boldsymbol{S}_i \boldsymbol{S}_j, \qquad (1)$$

where $J_{ij}$ is the exchange interaction integral for $i$-th и $j$-th spins, and the summation is performed over the whole spin lattice.

The energy of interaction between the spins and "random local anisotropy" type defects is

$$W_{def} = -K_0 \sum_l (\boldsymbol{S}_l \boldsymbol{n}_l)^2, \qquad (2)$$

where the summation is performed over defects randomly located in the lattice sites, $K_0 > 0$ is random anisotropy constant, and $\boldsymbol{n}_l$ is a unit vector prescribing the random easy axis direction.

Switching to the continuous distribution of the order parameter $\boldsymbol{\eta}$, we introduce the inhomogeneous exchange energy in the form [10]

$$\widetilde{W}_{ex} = -\frac{1}{2} \int d^d\boldsymbol{r}\, D\, \frac{\partial \eta^\perp}{\partial x_i} \frac{\partial \eta^\perp}{\partial x_i}, \qquad (3)$$

where $\boldsymbol{\eta} \sim \boldsymbol{S}_l / b^d$, $b$ is the interstitial distance, $D \sim J b^{2+d}$, $J$ is the exchange integral reflecting the interaction of the nearest neighbors, and $\boldsymbol{\eta}^\perp(\boldsymbol{r})$ is the order parameter component orthogonal to the direction of its mean quantity $\boldsymbol{\eta}_0$.

The energy of interaction with defects is

$$W_{def} = -b^d \int d^d\boldsymbol{r}\, K(\boldsymbol{r})(\boldsymbol{n}(\boldsymbol{r})\, \boldsymbol{\eta}(\boldsymbol{r}))^2, \qquad (4)$$

where

$$K(\boldsymbol{r}) = K_0 b^d \sum_l \delta(\boldsymbol{r} - \boldsymbol{r}_l). \qquad (5)$$

## 3. Imry-Ma disordered state

Let us reproduce the arguments by Imry and Ma [1] as applied to defects of the "random local anisotropy" type. In the case of substantially non-collinear distribution of local easy axes directions in the order parameter space, the defects with a particular direction of their easy axes dominate in a system volume with some characteristic dimension $L$ due to concentration fluctuations, and thus a certain averaged anisotropy arises, the corresponding constant is

$$K_L \sim K_0 \sqrt{c} \left(\frac{b}{L}\right)^{d/2}, \qquad (6)$$

where $c$ is the dimensionless concentration of defects (the number of defects per a unit cell). If the order parameter vector follows the space fluctuations of the easy axis directions, then it takes place a gain in the volume anisotropy energy density comparing to a homogeneous state. The addition to the volume energy density comprises the quantity of the following order

$$w_{fl} \sim \frac{-K_L S^2}{b^d} \sim -K_0 S^2 \sqrt{c} (bL)^{-d/2} \propto L^{-d/2}, \qquad (7)$$

where $S$ is the spin vector modulus.

Due to a subsequent inhomogeneity in the order parameter the volume density of the exchange energy increases by the value

$$w_{ex} \sim \frac{JS^2}{b^{d-2} L^2} \propto L^{-2}. \qquad (8)$$

Hence for the space dimension $d<4$ the long-wave (corresponding to big $L$) fluctuations of the order parameter direction become energetically favorable and the Imry-Ma disordered state arises. The optimum size $L^*$ corresponding to a minimum overall energy density $w = w_{fl} + w_{ex}$ equals

$$L^* \sim b \left(\frac{J^2}{cK_0^2}\right)^{\frac{1}{4-d}}. \qquad (9)$$

For the Imry-Ma state the addition to the volume density of the ordered state energy is

$$w \sim -\frac{K_0 S^2}{b^d} c^{\frac{2}{4-d}} \left(\frac{K_0}{J}\right)^{\frac{d}{4-d}}. \qquad (10)$$

In the case of perfectly isotropic distribution of easy axes directions in the order parameter space, the arguments presented above provoke no objections.

However for the anisotropic distribution of easy axes directions one has to take into account the defect induced anisotropy of the system. As opposed to the "random local field" type defects when the effective anisotropy arises in the random field second order term [5, 7], the "random local anisotropy" type defects induce the anisotropy to the first order term in the constant $K_0$. Let us consider sequentially the cases of the *X-Y* model and the Heisenberg model in three- and two-dimensional space.

## 4. X-Y model

Let us take a look at the anisotropic distribution of the easy axes directions of defects in the two-dimensional space of the order parameter in the form

$$\rho(\boldsymbol{n}) = A\big[n_x^2 + (1+\varepsilon)n_y^2\big], \tag{11}$$

where $\rho(\boldsymbol{n})$ is the probability density for the given distribution, $n_x$ and $n_y$ are vector $\boldsymbol{n}$ projections onto the axes of the Cartesian coordinate system, and $\varepsilon > 0$ is the degree of the distribution asymmetry.

The defect induced global effective anisotropy in the order parameter space (characterized by the constant $K_{eff}$) is defined by the difference between the anisotropy energies of the ordered states with the order parameters parallel to *y* and *x* axes respectively. In the limiting case of collinear local anisotropy axes of defects it takes its maximum equal to $K_{eff}^{max} = cK_0$.

By averaging over the distribution given by Eq. (11) one finds

$$\frac{K_{eff}}{K_{eff}^{max}} = \frac{3\varepsilon}{4(2+\varepsilon)}. \tag{12}$$

For the case of the *X-Y* model, this asymmetry of the easy axes distribution induces in the system the global anisotropy of the "easy axis" type.

If the anisotropy constant $K_{eff}$ magnitude exceeds the critical value, then the long range order state remains the ground one. Indeed, to follow the space fluctuations of the easy axis direction the order parameter is led to deviate from the global easy axis. This circumstance causes the anisotropy volume energy growth by the value of the order of $K_{eff}S^2/b^d$. When this value exceeds the modulus of the quantity given by Eq. (10), the Imry-Ma disordered state ceases to be energetically favorable. From this results the necessary condition for the existence of the Imry-Ma state

$$\varepsilon < (1 \div 10)c^{\frac{d-2}{4-d}}. \tag{13}$$

For $d=3$, $c\sim 10^{-2}$, and $K_0/J \sim 10^{-2}$ one obtains the inequality for $\varepsilon$ magnitude: $\varepsilon < 10^{-7}$. A formation of the isotropic distribution of easy axes with such a precision in a real crystal seems to be a stubborn problem. Of course, it is not inconceivable that metastable disordered states do exist. (For comparison, analogous calculations for defects of the "random local field" type at $c\sim 10^{-2}$ and the random field $h\sim 0,1J$ yield a restriction: $\varepsilon < 10^{-3}$).

In the two-dimensional space ($d=2$) the defect induced global anisotropy transforms the system to the class of Ising models [11] and causes the occurrence of the long range order at finite temperature. Similar physical reasoning provides the necessary condition for the existence of the Imry-Ma ground state in the two-dimensional space $\varepsilon < 10^{-1} \div 10^{-2}$. Such a condition can be fulfilled in real experiments.

## 5. Heisenberg model

For the three-dimensional space of the order parameter ($d=3$) we consider the anisotropic distribution of random easy axes directions in the form

$$\rho(\boldsymbol{n}) = A\left[n_x^2 + n_y^2 + (1+\varepsilon)n_z^2\right]. \tag{14}$$

The effective anisotropy constant in the order parameter space $K_{eff}$ is determined by the difference of the anisotropy energies of the ordered states with the order parameter being parallel and perpendicular to the $z$ axis respectively. The elementary calculation brings

$$\frac{K_{eff}}{K_{eff}^{max}} = \frac{2\varepsilon}{5(3+\varepsilon)} . \qquad (15)$$

At $\varepsilon > 0$ the global anisotropy of the "easy axis" type arises in the system, with a consequent transformation of the system to the class of Ising models.

The necessary condition for the existence of the Imry-Ma ground state in the space dimensions $d=3$ and $d=2$ gives the same limitations on the anisotropy degree as in the case of the X-Y model ($\varepsilon < 10^{-7}$ and $\varepsilon < 10^{-1}$ respectively).

In the case $-1 < \varepsilon < 0$, the global anisotropy of the "easy plane" type is induced by the defects. The limiting case corresponds to the coplanar and isotropic in the *xy* plane distribution of the anisotropy easy axes.

The effect of such an anisotropy transforms the system to the class of *X-Y* models. The question on the occurrence of the long range order or the Imry-Ma state at the presence of asymmetrical distribution of easy axes in the "easy plane" *xy* can be answered by vectors $\boldsymbol{n}_l$ projecting onto this plane and considering the problem in accordance with the approach proposed in Section 4.

## 6. Conclusion

As shown above, the anisotropic distribution of random local anisotropy axes induced by defects initiates the global anisotropy in the order parameter space.

If the constant of the "easy axis" type global anisotropy exceeds some threshold value, then the long range order in the system with initial *O(n)* symmetry and the space dimensionality 2<*d*<4 holds and the Imry-Ma disordered state does not arise.

For initiation of the Imry-Ma type ground state in the system it is necessary that the anisotropy degree of the distribution of random local easy axes orientations does not exceed the value of the order of $10^{-7}$ in the three-dimensional system and the value of the order of $10^{-1}$ in the two-dimensional one.